\documentclass[aps,pra,twocolumn,reprint,superscriptaddress,nofootinbib]{revtex4-2}

\usepackage[svgnames,table,xcdraw]{xcolor}
\usepackage{amsmath}
\usepackage{microtype}
\usepackage{amssymb}
\usepackage{amsmath}
\usepackage{braket}
\usepackage{graphicx}
\usepackage{booktabs}
\usepackage{bm}
\usepackage{dsfont}
\usepackage{hyperref}
\usepackage[caption=false]{subfig}
\usepackage{amsthm}

\usepackage{xfrac}
\usepackage{enumitem}
\usepackage{orcidlink}
\usepackage{graphicx}
\usepackage{float}
\usepackage{threeparttable}
\usepackage{makecell}

\hypersetup{colorlinks,linkcolor={DarkBlue},citecolor={DarkBlue},urlcolor={DarkBlue}}

\let\originalleft\left
\let\originalright\right
\renewcommand{\left}{\mathopen{}\mathclose\bgroup\originalleft}
\renewcommand{\right}{\aftergroup\egroup\originalright}

\usepackage{nag}
\usepackage{graphicx}
\usepackage{amsthm}
\usepackage{tabulary}
\usepackage{qcircuit}
\usepackage{mathtools}
\usepackage{bm}
\usepackage{braket}
\usepackage{datetime}
\usepackage{stackrel}
\usepackage{dsfont}
\usepackage{qcircuit}
\usepackage[english]{babel}
\usepackage{units}

\usepackage{tikz}
\usetikzlibrary{backgrounds,decorations.pathreplacing}

\usepackage{chngcntr}
\counterwithout{equation}{section}

\usepackage[normalem]{ulem}
\usepackage{slashed}
\usepackage{soul}

\usepackage{xcolor}
\usepackage{amssymb}
\usepackage{amsmath}
\usepackage{amsthm}
\usepackage{braket}
\usepackage{mathdots}

\usepackage{makeidx}
\makeindex

\usepackage{soul}
\usepackage{xargs}
\newcommandx{\cmnote}[2][1=]{\linespread{1.0}\todo[linecolor=red,backgroundcolor=red!25,bordercolor=red,#1]{#2}}

\let\underline\ul

\allowdisplaybreaks[4]

 \index{} \index{} \index{} \index{} \index{} \index{} \index{} \index{}
\makeatletter

\newcommand{\ringplus}{\mathbin{\text{\@ringplus}}}

\newcommand{\@ringplus}{%
  \ooalign{\hidewidth\raise1.3ex\hbox{\tiny$\circ$}\hidewidth\cr$\m@th+$\cr}%
}

\newcommand{\ringminus}{\mathbin{\text{\@ringminus}}}

\newcommand{\@ringminus}{%
  \ooalign{\hidewidth\raise0.9ex\hbox{\tiny$\circ$}\hidewidth\cr$\m@th-$\cr}%
}
\makeatother
 \index{} \index{} \index{} \index{} \index{} \index{} \index{} \index{}

\DeclareFontFamily{U}{wncy}{}
\DeclareFontShape{U}{wncy}{m}{n}{<->wncyr10}{}
\DeclareSymbolFont{mcy}{U}{wncy}{m}{n}
\DeclareMathSymbol{\Sh}{\mathord}{mcy}{"58}

\renewcommand{\vec}[1]{\bm{#1}}

 \index{} \index{} \index{} \index{} \index{} \index{} \index{} \index{}
 \index{} \index{} \index{} \index{} \index{} \index{} \index{} \index{}

 \index{} \index{} \index{} \index{} \index{} \index{} \index{} \index{}
 \index{} \index{} \index{} \index{} \index{} \index{} \index{} \index{}
\usepackage{graphicx}

\usepackage{graphicx}
\usepackage{multirow}
\usepackage{hhline}

\xyoption{color}
\xyoption{line}

\newcommandx*\bsbal[3][1=black, 3=->]{\ar @[#1]@{#3} [#2,0] \qw}

\newcommandx*\varbs[5][1=black, 3=\theta,4=0.5,5=->]{\ar @[#1]@{#5}^(#4){#3} [#2,0] \qw}

\newcommandx*\lblline[3][3=0.5]{\ar @{-}^(#3){#1} [#2,0]}
\newcommandx*\ctrlg[3][3=0.5]{ \raisebox{-3pt}{$\bullet$}  \ar @{-}^(#3){#1} [#2,0] \qw }
\newcommandx*\ctrlog[2]{\controlo \ar @{-}^{#1} [#2,0] \qw}
\newcommandx*\ctrlodash[1]{\controlo \ar @{-} [#1,0] \ar @[black]@{.} [0,-1]}

\begin{document}


\title{%
  \texorpdfstring
  {First-Principles Optical Descriptors \\ and Hybrid Classical–Quantum Classification of Er-Doped CaF$_2$}
  {First-Principles Optical Descriptors and Hybrid Classical–Quantum Classification of Er-Doped CaF$_2$}
}

\def \affdavid {Divisi\'{o}n de Ciencias e Ingenier\'{i}as Campus Le\'{o}n, Universidad de Guanajuato, Loma del Bosque 103, Le\'{o}n, Guanajuato 37150, M\'{e}xico}

\def \affkeremee {Department of Electrical and Electronics Engineering, Bogazici University, Istanbul, 34342, Turkey}
\def \affkeremphys {Department of Physics, Bogazici University, Istanbul, 34342, Turkey}

\def \affkrishan {Centre for Development of Advanced Computing, New Delhi, India}

\def \affragu {Independent Quantum Machine Learning Researcher, Coventry, United Kingdom}

\def \affkhizar {Department of Electrical Engineering, FAST-NUCES, Islamabad, Pakistan}

\def \affAlirezaIIT {Department of Physics, Illinois Institute of Technology, Chicago, IL 60616, USA}
\def \affGatech {College of Computing, Georgia Institute of Technology, Atlanta, GA 30332, USA}

\author{David Angel Alba Bonilla}
\affiliation{\affdavid}

\author{Kerem Yurtseven}
\affiliation{\affkeremee}
\affiliation{\affkeremphys}

\author{Krishan Sharma}
\affiliation{\affkrishan}

\author{Ragunath Chandrasekharan}
\affiliation{\affragu}

\author{Muhammad Khizar}
\affiliation{\affkhizar}

\author{Alireza Alipour\,\orcidlink{0000-0002-4549-330X}}
\affiliation{\affAlirezaIIT}

\author{Dennis Delali Kwesi Wayo}
\affiliation{\affGatech}

\date{\today}

\begin{abstract}
We present a physics-informed classical–quantum machine learning framework for discriminating pristine CaF$_2$ from Er-doped CaF$_2$ using first-principles optical descriptors. Finite Ca$_8$F$_{16}$ and Ca$_7$ErF$_{16}$ clusters were constructed from the fluorite structure ($a=5.46$~\AA) and treated using density functional theory (DFT) and linear-response time-dependent DFT (LR-TDDFT) within the GPAW code. Geometry optimization was performed in LCAO mode with a DZP basis and PBE exchange–correlation functional, followed by real-space finite-difference ground-state calculations with grid spacing $h=0.30$~\AA\ and $N_{\text{bands}}=N_{\text{occ}}+20$. Optical excitations up to 10~eV were obtained via the Casida formalism and converted into continuous absorption spectra using Gaussian broadening ($\sigma=0.1$–0.2~eV). From 1,589 energy-resolved points per system, physically interpretable descriptors including transition energy $E$, extinction coefficient $\kappa$, and absorption coefficient $\alpha$ were extracted. A classical RBF-kernel support vector machine (SVM) achieves a test accuracy (ACC) of 0.983 and ROC–AUC of 0.999. Quantum support vector machines (QSVMs) evaluated on statevector and noisy simulators reach accuracies of 0.851 and 0.817, respectively, while execution on IBM quantum hardware yields a test-slice accuracy of 0.733 under finite-shot and decoherence constraints. A hybrid quantum neural network (QNN) with a 3-qubit feature map and depth-4 ansatz achieves a test accuracy of 0.93 and AUC of 0.96. Results here demonstrate that dopant-induced optical fingerprints form a robust, physically grounded feature space for benchmarking near-term quantum learning models against strong classical baselines.
\end{abstract}
\maketitle

\section{Introduction}
Photon upconversion is a nonlinear optical process in which the sequential absorption of two or more low-energy photons leads to the emission of a single higher-energy photon through intermediate electronic states \cite{Auzel2004,Boyd2008,wayo2025atomistic}. Rare-earth (lanthanide) ions play a central role in upconversion materials due to their partially filled electronic shells $4f$, which give rise to ladder-like energy level structures with long-lived metastable states that enable efficient stepwise excitation under low optical intensities \cite{Haase2011,Zhou2015}. Among these ions, Er$^{3+}$- and Yb$^{3+}$-doped systems are particularly prominent because of favorable energy-level matching and efficient energy transfer pathways \cite{Chen2015}. The efficiency of upconversion is strongly influenced by the host lattice, with low-phonon energy fluoride materials such as CaF$_2$ suppressing nonradiative relaxation and providing a chemically compatible environment for rare-earth substitution \cite{wang2009}. These properties have generated extensive interest in rare-earth upconversion materials for applications ranging from solid-state lasers to photovoltaics, bioimaging, and emerging photonic and quantum technologies \cite{Haase2011,Zhou2015}.

Despite the widespread use of rare-earth upconversion materials, a quantitative first-principles description of upconversion processes remains challenging. Conventional ground-state density functional theory (DFT) accurately captures structural and electronic properties \cite{wayo2025dft}, but is inherently limited to static observables and cannot describe optical excitations or multiphoton dynamics \cite{Sholl2022}. Time-dependent density functional theory (TDDFT) extends DFT to the time domain and enables the calculation of optical absorption spectra and electronic excitation energies \cite{Marques2012}, yet primarily probes linear-response excitations and does not directly capture nonlinear emission processes such as photon upconversion. In rare-earth–doped systems, this limitation is particularly pronounced, as experimentally observed visible upconversion emissions often originate from long-lived intermediate states and sequential energy transfer pathways that lie outside the scope of direct TDDFT absorption spectra \cite{Auzel2004,Zhou2015}. As a result, identifying physically meaningful signatures of upconversion requires going beyond direct spectral inspection and extracting robust descriptors that encode trends across materials and doping configurations. This challenge motivates the integration of first-principles spectroscopy with data-driven analysis methods, providing a systematic pathway to relate electronic excitation landscapes to photonic functionality.

First-principles optical spectra obtained from TDDFT calculations contain rich information about electronic excitations, yet their direct interpretation across different materials and doping configurations is often nontrivial. To enable systematic comparison, it is advantageous to construct compact spectral descriptors that encode physically relevant features such as peak positions, intensities, bandwidths, and energy separations \cite{Butler2018}. These descriptors provide a reduced representation of the excitation landscape while preserving sensitivity to the underlying trends of electronic structures. Data-driven approaches based on machine learning (ML) have proven effective in uncovering correlations between such descriptors and functional material properties, particularly in cases where explicit analytic models are unavailable or incomplete \cite{Schmidt2019,wayo2025ensembles}. In the context of photonic and optical materials, ML methods have been successfully applied to classify spectral responses, identify dominant transitions, and guide materials discovery workflows \cite{Raccuglia2016}.

Beyond classical ML, recent developments have explored the use of quantum machine learning (QML) algorithms, which take advantage of parameterized quantum circuits to process high-dimensional data in Hilbert spaces that may be difficult to access classically \cite{Biamonte2017}. Although no general quantum advantage has been established for materials modeling tasks, QML provides a complementary framework for investigating alternative feature embeddings and kernel constructions, particularly for structured spectral data derived from quantum mechanical simulations \cite{Schuld2022}. In this work, QML is not assumed to outperform classical approaches; instead, it is investigated as an exploratory tool to assess whether quantum-inspired representations can capture nontrivial relationships within TDDFT-derived spectral descriptors. This positioning enables a balanced comparison between classical and quantum-enhanced models while maintaining a physics-driven interpretation of the learned features.

In this work, we present a first-principles–driven framework for analyzing photon upconversion in rare-earth–doped materials: CaF$_2$ and CaF$_2$:Er. Our approach focuses on extracting physics-informed spectral descriptors from TDDFT optical absorption spectra of these systems, leveraging linear-response excitation information as a consistent and transferable basis for comparative analysis. These descriptors are then used as input features to benchmark classical machine learning against quantum machine learning models, enabling the distinction between pristine CaF$_2$ and Er-doped CaF$_2$.

The primary contributions of this study are threefold. First, we establish a systematic procedure for constructing spectral descriptors from TDDFT-calculated optical spectra, enabling meaningful comparison across materials and doping configurations. Second, we evaluated the effectiveness of classical machine learning models in identifying meaningful relationships between these descriptors and the corresponding materials. Third, we explore quantum machine learning models as an exploratory extension to assess whether quantum-enhanced representations provide complementary insights relative to classical approaches, without presupposing a performance advantage. 

The remainder of this paper is organized as follows. Section~\ref{sec:methodology} describes the computational and theoretical framework, including model systems, TDDFT calculations, optical spectrum construction, feature engineering, and the classical and quantum machine learning models employed for the classification of CaF$_2$ and Er-doped CaF$_2$. Section~\ref{sec:results} presents the results of the TDDFT calculations, optical spectra, feature engineering, and the benchmarking of classical and quantum machine learning approaches. Section~\ref{sec:discussion} discusses the results and compares them with those reported in other studies using QML for materials classification and regression tasks, as well as with results obtained on benchmark datasets. Finally, Section~\ref{sec:conclusion} summarizes the main findings and outlines directions for future work.

%
\section{Methodology} \label{sec:methodology}
%
\subsection{Electronic-Structure and Optical Response of CaF$_2$ and CaF$_2$:Er}
The electronic and optical properties of pristine CaF$_2$ and erbium-doped CaF$_2$ were investigated using a first-principles workflow combining density functional theory (DFT) and linear-response time-dependent density functional theory (LR-TDDFT), as implemented in the GPAW code~\cite{enkovaara2010electronic,mortensen2005real} within the Atomic Simulation Environment (ASE)~\cite{larsen2017atomic}. The methodology was designed to generate physically interpretable optical descriptors suitable for subsequent quantum and classical machine learning analysis.
\paragraph*{Cluster construction and doping model.}
Bulk CaF$_2$ was constructed in the fluorite structure using the experimental lattice constant $a = 5.46~\text{\AA}$~\cite{o2020crystal}. A finite cluster model was generated by repeating the primitive cell in a $2\times2\times2$ supercell configuration, yielding a stoichiometric Ca$_8$F$_{16}$ cluster (24 atoms). A vacuum padding of 6--10~\text{\AA} was applied in all directions to eliminate spurious interactions arising from periodic boundary conditions. Erbium doping was modeled via substitution of a single Ca atom with Er, resulting in a Ca$_7$ErF$_{16}$ cluster, corresponding to a dilute dopant regime appropriate for optical spectroscopy studies~\cite{moos1970spectroscopic,reisfeld1987optical}. The pristine CaF$_2$ cluster is defect-free, while the CaF$_2$:Er system contains a single, well-defined substitutional dopant, allowing isolation of impurity-induced electronic and optical effects without additional structural disorder.
\paragraph*{Geometry optimization and ground-state DFT.}
Initial structural relaxation was performed using the linear combination of atomic orbitals (LCAO) mode with a double-$\zeta$ polarized (DZP) basis set and the Perdew--Burke--Ernzerhof (PBE) exchange--correlation functional~\cite{perdew1996generalized}. To ensure robust self-consistent field (SCF) convergence for finite clusters, a Pulay-type density mixer with reduced mixing parameter ($\beta = 0.05$) and loose convergence thresholds was employed. Geometry optimization proceeded until the maximum residual force satisfied $|\mathbf{F}_i| < 0.15~\text{eV/\AA}$. Following relaxation, the electronic ground state was recomputed using a real-space finite-difference (FD) representation with grid spacing $h = 0.30~\text{\AA}$. The total number of Kohn--Sham states was chosen as shown in Equation~\ref{eqn:num_bands}
\begin{equation}
N_{\text{bands}} = N_{\text{occ}} + N_{\text{virt}},
\label{eqn:num_bands}
\end{equation}
where $N_{\text{virt}} = 20$ virtual states ensured convergence of unoccupied excitations. Electronic occupations were smoothed using a Fermi--Dirac distribution with width $k_B T = 0.03~\text{eV}$.
\paragraph*{Linear-response TDDFT and optical spectra.}
Excited-state properties were computed using linear-response TDDFT in the Casida formalism~\cite{casida1995time}. The interacting excitation energies $\Omega_I$ were obtained from the generalized eigenvalue problem (Equation~\ref{eq:eigenvalue_problem}),
\begin{equation}
\begin{aligned}
\sum_{j b}
\Big[
& \delta_{i j}\,\delta_{a b}\,(\varepsilon_a - \varepsilon_i)^2
\\
&+ 4 \sqrt{\varepsilon_a - \varepsilon_i}\;
K_{i a, j b}\;
\sqrt{\varepsilon_b - \varepsilon_j}
\Big]
F^{(I)}_{j b}
=
\Omega_I^2\, F^{(I)}_{i a},
\end{aligned}
\label{eq:eigenvalue_problem}
\end{equation}
where $\varepsilon_i$ and $\varepsilon_a$ denote occupied and unoccupied Kohn--Sham eigenvalues, respectively, and $K_{ia,jb}$ is the electron--hole interaction kernel. Oscillator strengths were evaluated as shown in Equation~\ref{eq:osc_strength}
\begin{equation}
f_I = \frac{2}{3} \Omega_I \sum_{\alpha=x,y,z} 
\left| \langle \Psi_0 | \hat{r}_\alpha | \Psi_I \rangle \right|^2,
\label{eq:osc_strength}
\end{equation}
providing a direct measure of optical transition intensities~\cite{onida2002electronic}. Excitations ranging up to 10~eV were retained.
\paragraph*{Spectral broadening and data generation.}
Discrete excitation spectra were transformed into continuous absorption profiles by Gaussian broadening (Equation~\ref{eq:gaussian_broadening}),
\begin{equation}
A(E) = \sum_I f_I 
\exp\left[-\frac{(E - \Omega_I)^2}{2\sigma^2}\right],
\label{eq:gaussian_broadening}
\end{equation}
with $\sigma$ controlling the effective linewidth. The resulting spectra, along with derived optical quantities such as the real and imaginary dielectric functions $(\varepsilon_1, \varepsilon_2)$, refractive index $n$, extinction coefficient $\kappa$, and absorption coefficient $\alpha$, formed the physically grounded descriptor set used in downstream machine learning and quantum feature mapping analyses.
%
\subsection{Feature Engineering and Physical Descriptor Selection}
We transform all input variables using the Box--Cox power transformation to mitigate skewness, improve numerical stability across both classical and quantum learning models, and make the data more Gaussian-like. We employ the scikit-learn implementation of this transformation, as defined in Equation~\ref{eq:box_cox}
\begin{equation}
z^{(\lambda)} =
\begin{cases}
\frac{z^\lambda - 1}{\lambda},  & \lambda \neq 0, \\
\ln(z), & \lambda = 0,
\end{cases}
\label{eq:box_cox}
\end{equation}
where $z$ denotes the original input variable, $z^{(\lambda)}$ represents the transformed value, and $\lambda$ is the transformation parameter that controls the power applied to the data. The parameter $\lambda$ is estimated from the training data by maximizing the log-likelihood under a normal linear model for the transformed observations~\cite{Box2018}. 

\begin{figure}[H]
\centering
\includegraphics[width=\linewidth]{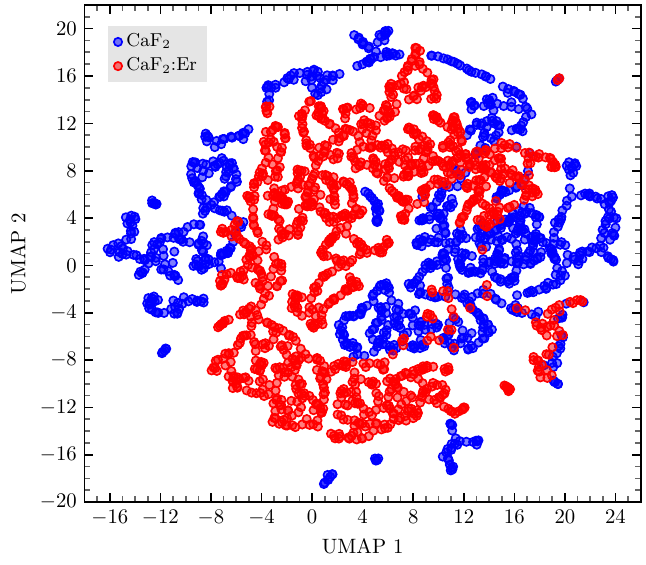}
\caption{UMAP projection of the three selected physical descriptors for CaF\(_2\) and CaF\(_2\):Er samples from the training set. The two-dimensional embedding is used exclusively for visualization purposes.}
\label{fig:umap_projection}
\end{figure}

To mitigate redundancy and form a set instead of an unordered sequence and identify the most discriminative physical descriptors, a linear support vector machine is employed to rank features based on the absolute magnitude of the learned weight coefficients. Based on this ranking, the three most relevant features selected for subsequent model training and evaluation are the absorption coefficient $\alpha$, the extinction coefficient $\kappa$, and the transition energy $E$. Uniform Manifold Approximation and Projection (UMAP) was applied exclusively for visualization purposes to explore the low-dimensional structure of the training set. The two-dimensional UMAP representation of the three selected physical descriptors is shown in Figure~\ref{fig:umap_projection}. Despite the inevitable information loss associated with dimensionality reduction, the projected data preserve a coherent structure and reveal distinct local organizations for CaF\(_2\) and CaF\(_2\):Er samples.
%
\subsection{Classical Machine Learning Baseline: Support Vector Machines for nonlinear separation}

Support Vector Machines (SVMs) were formally introduced in 1995~\cite{Cortes1995} and have since been successfully applied to a wide range of problems, including text categorization, image analysis, speech recognition, time-series forecasting, and information security. SVMs can be used for both classification and regression tasks; they are based on Statistical Learning Theory (SLT) and Structural Risk Minimization (SRM). Given an input training set $T=\{(\vec{x_1},y_1), \dotsb, (\vec{x_l},y_l)\}$, where $\vec{x_i} \in \mathbb{R}^n$, $y_i \in \{-1,1\}$, $i = 1, \dotsb, l$, the SVM method for classification is grounded on the following primal optimization problem (Equation~\ref{eq:svm_primal_problem}),
\begin{equation}
\begin{aligned}
\min_{\vec{w}, b, \vec{\xi} } \quad
& \frac{1}{2} \lVert \vec{w} \rVert^2 + C \sum_{i=1}^{l} \xi_i ,\\
\text{s.t.} \quad
& y_i(\vec{w} \cdot \Phi(\vec{x_i}) + b) \geq 1 - \xi_i , \\
& \xi_i \geq 0 , \quad i = 1, \dotsb, l ,
\end{aligned}
\label{eq:svm_primal_problem}
\end{equation}
where $C$ is the penalty parameter, $\Phi : \mathbb{R}^n \rightarrow \mathcal{H}$ maps the data into a Hilbert space, $b \in \mathbb{R}$ is the bias term, and $\xi_i$ are slack variables. From this formulation, the corresponding dual problem can be derived as shown in Equation~\ref{eq:svm_dual_problem}
\begin{equation}
\begin{aligned}
\min_{\vec{\alpha}} \quad
& \frac{1}{2} \sum_{i=1}^{l} \sum_{j=1}^{l}
y_i y_j \alpha_i \alpha_j K(\vec{x_i}, \vec{x_j})
- \sum_{j=1}^{l} \alpha_j ,\\
\text{s.t.} \quad
& \sum_{i=1}^{l} y_i \alpha_i = 0 , \\
& 0 \leq \alpha_i \leq C \, , \quad i = 1, \dotsb, l .
\end{aligned}
\label{eq:svm_dual_problem}
\end{equation}

Here, $K(\vec{x_i}, \vec{x_j})=\Phi(\vec{x_i})\cdot\Phi(\vec{x_j})$ is the kernel function, and $\vec{\alpha}=(\alpha_1, \dotsb, \alpha_l)$ denotes the vector of Lagrange multipliers. After solving this dual problem, the bias term $b$ can be obtained by selecting a Lagrange multiplier $\alpha_j \in (0, C)$, and evaluating Equation~\ref{eq:b_equation}
\begin{equation}
b = y_j-\sum_{i=1}^{l}y_i\alpha_iK(\vec{x_i}, \vec{x_j}).
\label{eq:b_equation}
\end{equation}

Finally, the decision function is given by Equation~\ref{eq:svm_decision_func}
\begin{equation}
f(\vec{x}) = \operatorname{sgn}\left( \sum_{i=1}^{l}y_i\alpha_iK(\vec{x_i}, \vec{x}) + b\right),
\label{eq:svm_decision_func}
\end{equation}
where $\operatorname{sgn}(\cdot)$ is a sign function defined in Equation~\ref{eq:sign_function}
\begin{equation}
\operatorname{sgn}(a) =
\begin{cases}
1,  & a \geq 0, \\
-1, & a < 0.
\end{cases}
\label{eq:sign_function}
\end{equation}

We employ the implementation of the Support Vector Classifier (SVC) provided by scikit-learn, which is based on LIBSVM~\cite{Chang2011}. The kernel used is the Radial Basis Function (RBF), as shown in Equation~\ref{eq:kernel_rbf}
\begin{equation}
K(\vec{x}, \vec{y}) = e^{-\gamma \lVert \vec{x} - \vec{y} \rVert^2}.
\label{eq:kernel_rbf}
\end{equation}

We use the default parameters of the RBF model, which are $C=1$ and $\gamma$ given by Equation~\ref{eq:gamma_equation}
\begin{equation}
\gamma = \frac{1}{n_{features} \operatorname{Var}(X)},
\label{eq:gamma_equation}
\end{equation}
where $n_{features}$ is the number of features and $\operatorname{Var}(X)$ denotes the variance of the input data.
%
\subsection{Quantum Support Vector Machine on Ideal and Noisy Simulators}
The quantum version of SVM was initially proposed in~\cite{Rebentrost2014}, but this formulation requires the input data to be provided in coherent superposition states. Therefore, when the data are generated on a classical computer, the efficient application of this approach becomes impractical.
A different approach was proposed in~\cite{Havlicek2019}, where the data are provided purely classically, while the feature space is mapped by the quantum state $\Phi: \mathbb{R}^n \rightarrow \ket{\Phi(\vec{x})} \bra{\Phi(\vec{x})}$. For this work, we selected the \texttt{ZZFeatureMap} as the quantum feature map (Equation~\ref{eq:zzfeature_map})
\begin{equation}
\begin{aligned}
\mathcal{U}_{\Phi(\vec{x})}
&=
\left(
\exp\!\left(
i \sum_{jk} \phi_{\{j,k\}}(\vec{x})\, Z_j \otimes Z_k
\right)
\right. \\
&\quad \left.
\times
\exp\!\left(
i \sum_{j} \phi_{\{j\}}(\vec{x})\, Z_j
\right)
H^{\otimes n}
\right)^{d}.
\label{eq:zzfeature_map}
\end{aligned}
\end{equation}

Here, $Z$ denotes the Pauli-$Z$ operator, $n$ denotes the number of qubits (corresponding to the number of features), $d$ is the number of repetitions, and $\phi_S(\vec{x})$ is defined in Equation~\ref{eq:phi_feature_map}
\begin{equation}
\phi_S(\vec{x}) =
\begin{cases}
x_i, & S = \{i\} ,\\
(\pi - x_i)(\pi - x_j), & S = \{i,j\} .
\end{cases}
\label{eq:phi_feature_map}
\end{equation}

The quantum kernel is represented as shown in Equation~\ref{eq:kernel_quantum}
\begin{equation}
\begin{aligned}
K(\vec{x}, \vec{y})
&= \lvert 
\left\langle \Phi(\vec{x}) \middle| \Phi(\vec{y}) \right\rangle 
\rvert^2 \\
&= \lvert \bra{0^{\otimes n}} \mathcal{U}_{\Phi(\vec{x})}^\dagger \mathcal{U}_{\Phi(\vec{y})} \ket{0^{\otimes n}}\rvert^2.
\end{aligned}
\label{eq:kernel_quantum}
\end{equation}

\begin{figure*}
\centering
\includegraphics[width=0.95\linewidth]{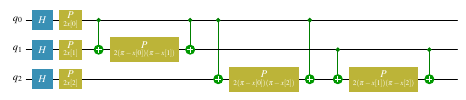}
\caption{Quantum circuit diagram of the feature map $\mathcal{U}_{\Phi(\vec{x})}$ (\texttt{ZZFeatureMap}) for 3 qubits, used to encode classical data into a quantum feature space for the construction of the QSVM kernel.}
\label{fig:zz_feature_map}
\end{figure*}

We employ the Qiskit library to construct the quantum circuits implementing the \texttt{ZZFeatureMap} (see Figure~\ref{fig:zz_feature_map}) and to perform quantum kernel evaluation using the \texttt{FidelityQuantumKernel} module. 
The feature map is implemented with 3 qubits and a single repetition. The resulting kernel matrix, obtained by evaluating the quantum kernel for all pairs of data points, is subsequently used to train a classical SVC. The quantum kernel was evaluated on two simulators: a statevector (SV) simulator and a QASM-style (shot-based) simulator under a depolarizing noise model with an error probability of $p=0.05$ on single-qubit and two-qubit gates.
%
\subsection{Quantum Kernel Evaluation on IBM Quantum Hardware}
\begin{figure}[H]
\centering
\includegraphics[width=0.80\linewidth]{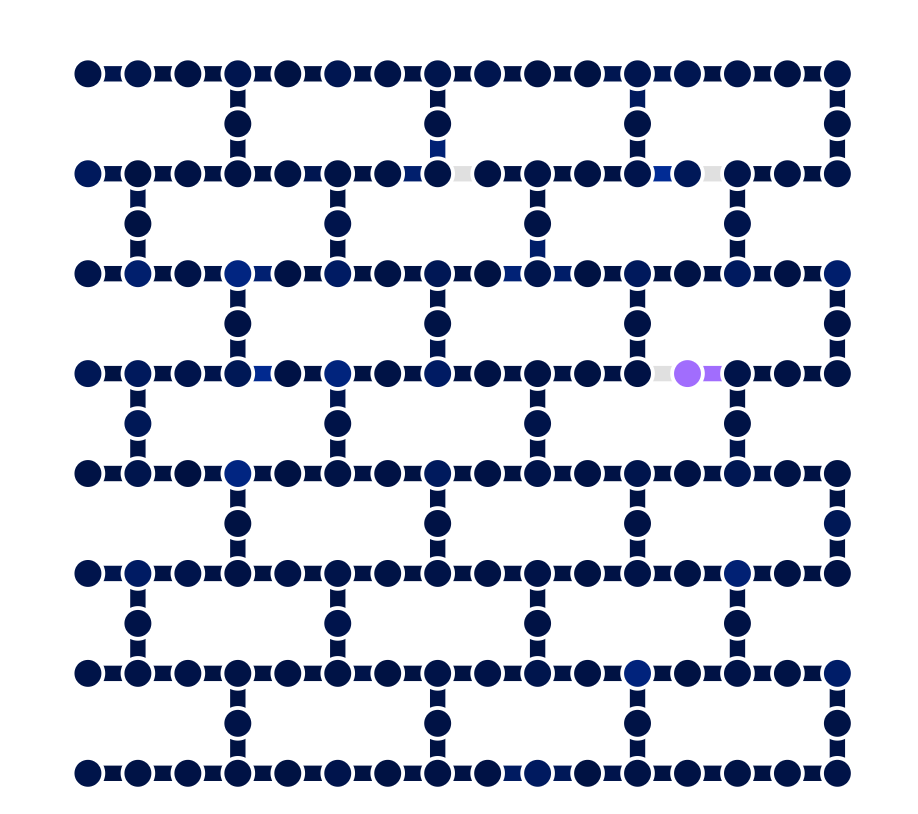}
\caption{Layout of the qubit connectivity for the 156-qubit IBM Fez device. Accessed from the IBM Quantum Platform on January 19, 2026.}
\label{fig:ibm_fez}
\end{figure}

To assess real-world feasibility, quantum kernel evaluation was performed on IBM Quantum hardware using the \texttt{ibm\_fez} superconducting processor (see Figure~\ref{fig:ibm_fez}), which is accessed via the IBM Quantum Platform.
The quantum feature map and kernel construction follow the same approach as in the simulator-based model, but with reduced datasets for training (30 samples) and testing (15 samples). All quantum circuits were transpiled using the Qiskit module \texttt{generate\_preset\_pass\_manager} with \texttt{optimization\_level = 2}, which performs logical-to-physical qubit mapping and optimizes circuit depth and gate count according to the connectivity and operational constraints of the target IBM Quantum hardware.
%
\subsection{Quantum Neural Networks}\label{subsec:method_qnn}
\begin{figure*}
\centering
\includegraphics[width=0.95\linewidth]{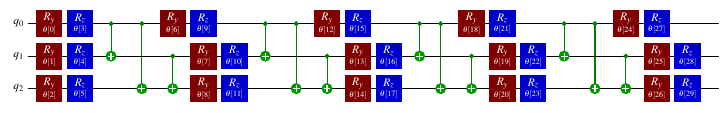}
\caption{Schematic of the parameterized \texttt{TwoLocal} ansatz $U(\vec{\theta})$ serving as the variational block of the QNN. The circuit consists of $L=4$ repetitions of rotation layers (containing parameterized $R_y$ and $R_z$ gates) and entanglement layers (using all-to-all CNOT gates). This structure provides the necessary expressibility for the quantum processor to learn the complex decision boundary between pristine and Er-doped spectral signatures.}
\label{fig:qnn_ansatz}
\end{figure*}

While the QSVM relies on a fixed, pre-computed kernel to measure similarity in Hilbert space, the Variational Quantum Classifier (VQC) employs a flexible, parameterized quantum circuit (PQC) that learns the optimal decision boundary directly~\cite{Cerezo2021}. In this framework, the quantum processor functions as a trainable layer within a larger hybrid quantum-classical optimization loop. We implemented a Quantum Neural Network (QNN) using the \texttt{PyTorch} and \texttt{Qiskit} interface~\cite{Benedetti2019}. The architecture consists of three distinct stages: state preparation (feature map), variational processing (ansatz), and measurement (readout). First, the classical input vector $\vec{x} \in \mathbb{R}^n$ (where $n=3$, corresponding to $\alpha$, $\kappa$, and $E$) is encoded into the quantum state $|\Phi(\vec{x})\rangle$ using the \texttt{ZZFeatureMap} same as the QSVM. This map explicitly utilizes $Z$ and $ZZ$ interactions to generate the data-dependent state expressed as in Equation~\ref{eq:zzfeature_map}. Feature mapping circuit is illustrated in Figure~\ref{fig:zz_feature_map}. Second, a parameterized unitary ansatz $U(\vec{\theta})$ is applied to the encoded state. To balance circuit complexity with representational power, we selected a \texttt{TwoLocal} ansatz composed of parameterized single-qubit rotations ($R_y$ and $R_z$) and two-qubit entangling gates ($CX$). The circuit is described in Equation~\ref{eq:two_local_ansatz}
\begin{equation}
U(\vec{\theta}) = \prod_{l=1}^{L} \left( U_{\text{ent}} \prod_{j=1}^{n} R_z(\theta_{j,l}^z) R_y(\theta_{j,l}^y) \right),
\label{eq:two_local_ansatz}
\end{equation}
where $L$ denotes the number of repetitions (layers) and $U_{\text{ent}}$ represents the full entanglement block using CNOT gates. To enhance the expressibility of the model relative to the shallow kernel used in the QSVM, we set the circuit depth to $L=4$ ($\texttt{reps}=4$)~\cite{Sim2019}. Ansatz circuit is illustrated in Figure~\ref{fig:qnn_ansatz}
Finally, we used the \texttt{SamplerQNN} primitive backed by the \texttt{AerSampler} in state vector mode ($\text{shots}=\text{None}$), ensuring deterministic gradient calculations via the parameter-shift rule or backpropagation \cite{Schuld2019}. The quantum output is not a simple expectation value; instead, we apply a parity function $f(x)$ to the measured bit strings to map the quantum states to class probabilities (Equation~\ref{eq:parity_func}),
\begin{equation}
f(x) = \text{popcount}(x) \pmod 2,
\label{eq:parity_func}
\end{equation}
where $\text{popcount}(x)$ counts the number of set bits in the bitstring. This maps the $2^n$ basis states to a binary distribution over the two classes~\cite{Farhi2018}.

\begin{figure*}
\centering\includegraphics[width=0.95\linewidth]{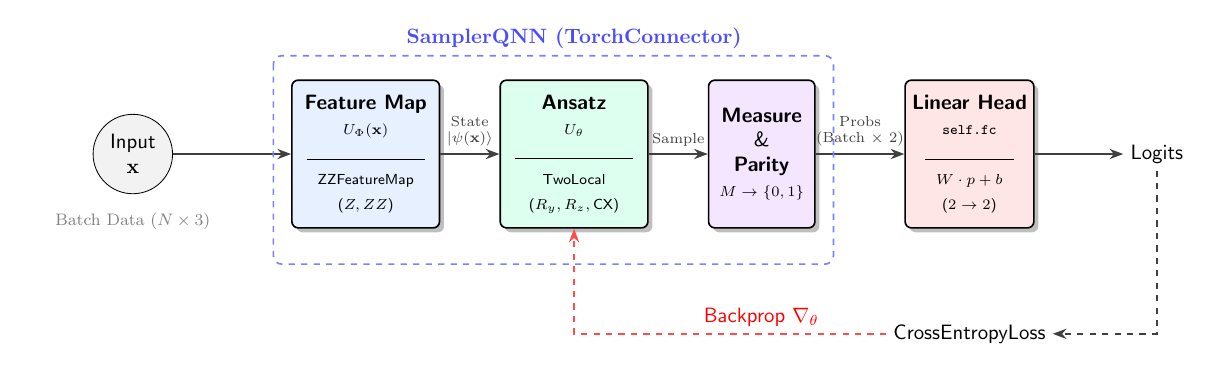}
\caption{High-level architecture of the hybrid quantum-classical neural network. The pipeline proceeds in four stages: (1) Encoding, where classical spectral data is mapped to the quantum processor; (2) Variational Processing, where the ansatz evolves the state based on trainable parameters $\vec{\theta}$; (3) Measurement, where bitstring parity is calculated to derive class probabilities; and (4) Classical Post-Processing, where a linear layer projects probabilities into logits. The entire model is optimized end-to-end using the Adam optimizer to minimize the Cross-Entropy loss.}
\label{fig:hybrid_qnn_arch}
\end{figure*}

The output of the QNN layer (class probabilities) is passed to a fully connected classical linear layer ($\texttt{nn.Linear}$) to project the features into logits. This hybrid structure allows the network to fine-tune the decision boundary beyond the constraints of the quantum Hilbert space alone. The model parameters $\vec{\theta}$ were optimized to minimize the categorical Cross-Entropy Loss, $\mathcal{L}(\vec{\theta})$, which quantifies the divergence between the predicted logits and the true class labels~\cite{Goodfellow2016}. Given a batch of $B$ samples, the loss function is defined in Equation~\ref{eq:qnn_loss}
\begin{equation}
\mathcal{L}(\vec{\theta}) = - \frac{1}{B} \sum_{i=1}^{B} \sum_{c=1}^{C} y_{i,c} \log \left( \frac{\exp(z_{i,c})}{\sum_{j=1}^{C} \exp(z_{i,j})} \right),
\label{eq:qnn_loss}
\end{equation}
where $y_{i,c}$ is the binary indicator (0 or 1) for class $c$, and $z_{i,c}$ represents the output logit for sample $i$ and class $c$ generated by the classical linear head. Minimization was performed using the \texttt{Adam} optimizer~\cite{Kingma2014}, which adapts the learning rate for each parameter based on estimates of the first and second moments of the gradients. The update rules at time step $t$ are given by Equations~\ref{eq:adam_update}
\begin{subequations}\label{eq:adam_update}
\begin{align}
m_t &= \beta_1 m_{t-1} + (1 - \beta_1) g_t, \\
v_t &= \beta_2 v_{t-1} + (1 - \beta_2) g_t^2, \\
\hat{m}_t &= \frac{m_t}{1 - \beta_1^t}, \quad \hat{v}_t = \frac{v_t}{1 - \beta_2^t}, \\
\theta_t &= \theta_{t-1} - \alpha \frac{\hat{m}_t}{\sqrt{\hat{v}_t} + \epsilon},
\end{align}
\end{subequations}
where $g_t = \nabla_{\vec{\theta}} \mathcal{L}_t(\vec{\theta}_{t-1})$ is the gradient, $\alpha$ is the learning rate, and $\beta_1, \beta_2$ are the exponential decay rates. We utilized the AMSGrad variant (`amsgrad=True`)~\cite{Reddi2018} to ensure convergence by using the maximum of past squared gradients $\hat{v}_t = \max(\hat{v}_{t-1}, v_t)$. To effectively navigate the optimization landscape and mitigate oscillations around local minima, we implemented a dynamic learning rate scheduler, \texttt{ReduceLROnPlateau}~\cite{Bengio2012}. The learning rate $\alpha$ is updated according to the validation loss $\mathcal{L}_{val}$ (Equation~\ref{eq:learning_rate}),
\begin{equation}
\alpha_{k+1} = 
\begin{cases} 
\gamma \alpha_k, & \text{if } \mathcal{L}_{val}^{(k)} \text{no improves for } P_{LR}, \\
\alpha_k, & \text{otherwise},
\end{cases}
\label{eq:learning_rate}
\end{equation}
where $\gamma = 0.5$ is the reduction factor and the patience is set to $P_{LR} = 2$ epochs. Finally, to prevent overfitting and ensure the model generalizes well to unseen spectral data, we employed a regularization strategy based on Early Stopping~\cite{Prechelt1998}. The training process terminates at epoch $e$ if the stopping condition in Equation~\ref{eq:qnn_stop_condition} is met,
\begin{equation}
\min_{t \in \{e-P_{ES}, \dots, e\}} \mathcal{L}_{val}(t) \geq \mathcal{L}^*_{val} - \delta,
\label{eq:qnn_stop_condition}
\end{equation}
where $P_{ES}=10$ is the patience parameter, $\delta$ is the minimum change threshold, and $\mathcal{L}^*_{val}$ represents the best validation loss recorded before the current window. We further imposed a hard constraint of $E_{min}=15$ minimum epochs to avoid premature stopping during the initial warm-up phase of the variational parameters. A full diagram of the hybrid quantum-classical neural network architecture is shown in Figure~\ref{fig:hybrid_qnn_arch}.
%
\section{Results} \label{sec:results}
\subsection{Electronic-Structure and Optical Response of CaF\(_2\) and CaF\(_2\):Er}
\begin{figure}[t]
\centering
\includegraphics[width=0.95\linewidth]{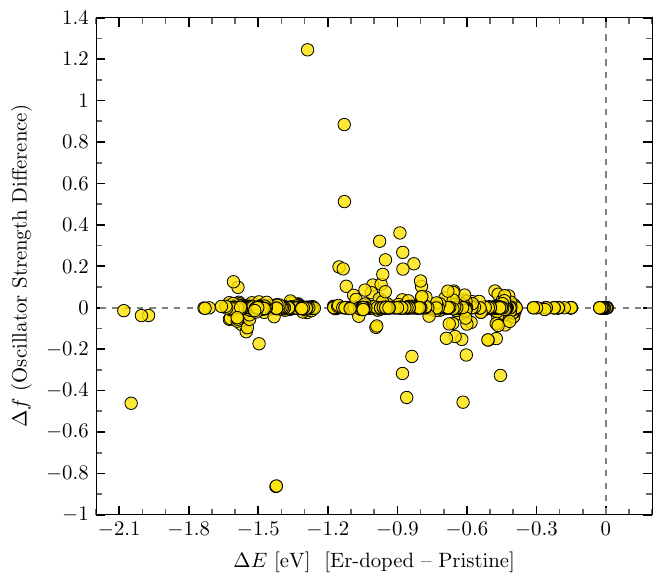}
\caption{Spectral difference map comparing CaF\(_2\):Er and pristine CaF\(_2\) optical transitions. Each point represents a single electronic transition, plotted as the change in transition energy $\Delta E$ (Er-doped minus pristine) versus the corresponding change in oscillator strength $\Delta f$. The distribution reveals systematic red-shifting of dominant transitions upon Er doping, accompanied by substantial redistribution of oscillator strength. The largest deviations cluster in the visible and near-UV energy range, highlighting the strong influence of localized Er\(^{3+}\) electronic states on the optical selection rule.}
\label{fig:spectral_difference}
\end{figure}

\begin{figure}[t]
\centering
\includegraphics[width=\linewidth]{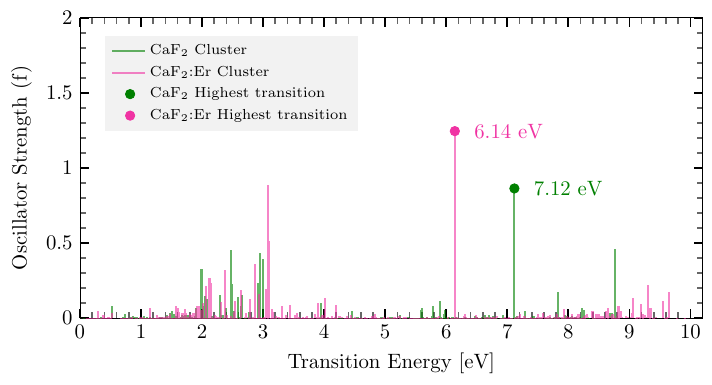}
\caption{Transition energy versus oscillator strength for pristine CaF\(_2\) and Er-doped CaF\(_2\) clusters. Pristine CaF\(_2\) exhibits sparse, high-energy dominant transitions characteristic of a wide-band-gap fluoride, with a prominent maximum near 7.12~eV. In contrast, CaF\(_2\):Er displays a dense manifold of mid-energy excitations and a shifted high-strength peak near 6.14~eV, reflecting dopant-induced modification of the excited-state electronic structure. Highlighted points indicate the strongest transitions for each system.}
\label{fig:transition_energy}
\end{figure}

\begin{figure*}[t]
\centering
\includegraphics[width=\linewidth]{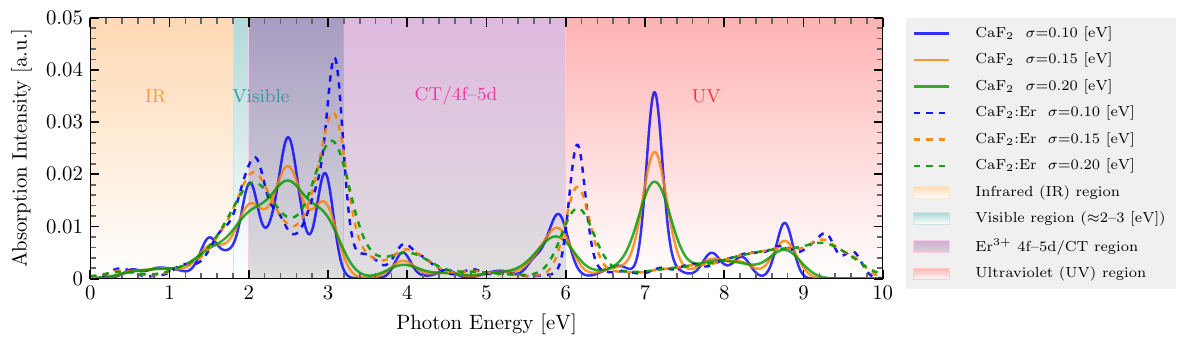}
\caption{Gaussian-broadened optical absorption spectra of pristine CaF\(_2\) and Er-doped CaF\(_2\) across infrared (IR), visible, charge-transfer (CT / Er\(^{3+}\) 4f–5d), and ultraviolet (UV) spectral regions. Solid lines correspond to pristine CaF\(_2\), while dashed lines represent CaF\(_2\):Er for Gaussian broadening widths $\sigma = 0.1$, 0.15, and 0.2~eV. Er incorporation induces pronounced absorption enhancement and spectral restructuring in the visible regime. At the same time, the persistence of these features across broadening widths indicates robustness against thermal and instrumental effects.}
\label{fig:broadened_spectra}
\end{figure*}

To quantify the dopant-induced perturbations, Figure~\ref{fig:spectral_difference} presents a spectral difference map between CaF\(_2\):Er and pristine CaF\(_2\) in terms of transition energy shifts ($\Delta E$) and oscillator strength changes ($\Delta f$). The distribution reveals a systematic red-shift of dominant transitions upon Er doping, with $\Delta E$ clustering between $-1.5$ and $-0.3$~eV. Simultaneously, significant positive and negative $\Delta f$ values emerge, confirming that Er incorporation redistributes oscillator strength rather than merely introducing weak perturbations. The largest deviations concentrate in the visible and near-UV regime, underscoring the role of Er-induced localized states in modifying optical selection rules.
Figure~\ref{fig:transition_energy} shows the transition-level landscape by plotting oscillator strength as a function of transition energy for both systems. While pristine CaF\(_2\) displays sparse, high-energy dominant transitions, CaF\(_2\):Er shows a dense manifold of mid-energy excitations with a pronounced high-strength peak near 6.14~eV, replacing the pristine CaF\(_2\) maximum at approximately 7.12~eV. This shift reflects a fundamental alteration of the excited-state electronic structure induced by Er doping.
Figure~\ref{fig:broadened_spectra} compares the Gaussian-broadened optical absorption spectra of pristine CaF\(_2\) and Er-doped CaF\(_2\) across infrared (IR), visible, charge-transfer (CT), and ultraviolet (UV) energy windows. For pristine CaF\(_2\), the absorption profile is dominated by high-energy interband transitions, with a pronounced UV peak at approximately 7.12~eV, consistent with its wide band-gap fluoride nature. The visible-region response remains weak and featureless, reflecting the absence of partially filled electronic states within the band gap. In contrast, Er incorporation induces substantial spectral restructuring. As shown in Figure~\ref{fig:broadened_spectra}, CaF\(_2\):Er exhibits enhanced absorption intensity and increased peak density in the visible region, with a maximum absorption peak at approximately 3.06~eV. This behavior is attributable to localized Er\(^{3+}\) 4f–4f and 4f–5d electronic transitions as well as ligand-to-metal charge-transfer states. Importantly, this enhancement persists across Gaussian broadening widths from $\sigma = 0.1$ to 0.2~eV, indicating that the observed spectral differences are robust against thermal and instrumental broadening effects rather than numerical artifacts.
These results do demonstrate that Er incorporation into CaF\(_2\) introduces structured, dopant-specific optical fingerprints spanning multiple spectral regions. The resulting nontrivial redistribution of transition energies and oscillator strengths provides a physically meaningful, high-dimensional feature space that is particularly well suited for quantum feature-map encoding in subsequent quantum machine learning analyses.
%
\subsection{Feature Engineering and Physical Descriptor Selection}
\label{subsec:feature_engineering}
\begin{figure}[H]
\centering
\includegraphics[width=\linewidth]{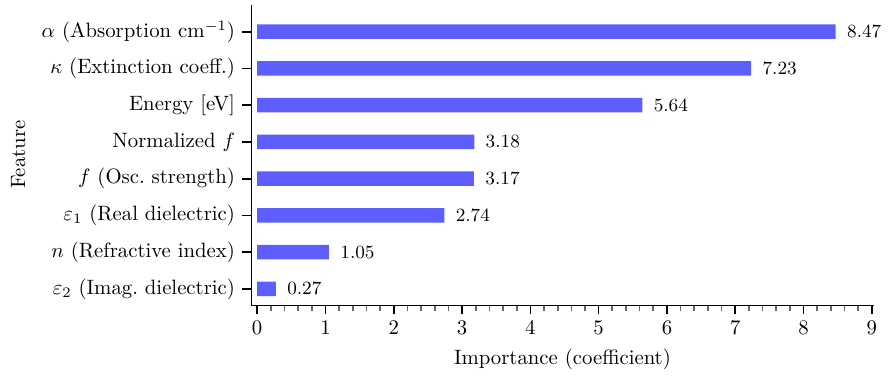}
\caption{Weights assigned to each feature when training a linear SVM model on the standardized dataset; these coefficients were used to select the top three features for subsequent analyses. The absorption coefficient $\alpha$ is the most influential feature, followed by the extinction coefficient $\kappa$ and transition energy $E$.}
\label{fig:svm_coeff_features}
\end{figure}

To bridge first-principles electronic-structure calculations with quantum machine learning (QML), a physically informed feature-engineering strategy was adopted for both pristine CaF$_2$ and Er-doped CaF$_2$:Er systems (see Table~\ref{tab:descriptor_samples} ). From the raw TDDFT-derived spectra, a compact set of optical and dielectric descriptors was extracted to capture complementary aspects of light–matter interaction. These descriptors include transition energies, oscillator strengths, normalized oscillator strengths, complex dielectric components, refractive indices, extinction coefficients, and absorption coefficients.
\begin{table}[H]
\centering
\caption{Representative optical descriptors extracted from first-principles simulations of CaF$_2$ and CaF$_2$:Er. The full dataset (1,589 energy points per system) is used for machine learning; selected rows are shown here to illustrate feature ranges and physical scales.}
\label{tab:descriptor_samples}
\begin{tabular}{cccccc}
\hline
$E$ (eV) 
& $\varepsilon_1$ 
& $\varepsilon_2$ 
& $n$ 
& $\kappa$ 
& $\alpha$ (cm$^{-1}$) \\
\hline
0.0063 & 4.4968 & $1.19\times10^{-6}$ & 2.1206 & $2.80\times10^{-7}$ & $2.51\times10^{-11}$ \\
0.0106 & 4.4947 & $2.12\times10^{-6}$ & 2.1201 & $5.00\times10^{-7}$ & $7.49\times10^{-11}$ \\
1.98   & 4.1125 & $3.41\times10^{-3}$ & 2.0287 & $8.42\times10^{-4}$ & $1.87\times10^{-5}$ \\
2.53   & 3.8841 & $1.87\times10^{-2}$ & 1.9724 & $4.74\times10^{-3}$ & $9.12\times10^{-4}$ \\
3.12   & 3.2148 & $4.62\times10^{-2}$ & 1.7943 & $1.29\times10^{-2}$ & $2.63\times10^{-3}$ \\
6.14   & 2.7846 & $1.25\times10^{-1}$ & 1.6129 & $3.92\times10^{-2}$ & $5.87\times10^{-3}$ \\
7.12   & 2.5039 & $8.74\times10^{-2}$ & 1.5681 & $2.91\times10^{-2}$ & $4.11\times10^{-3}$ \\
9.96   & 2.3412 & $1.11\times10^{-2}$ & 1.5310 & $3.62\times10^{-3}$ & $5.11\times10^{-4}$ \\
\hline
\end{tabular}
\end{table}
The descriptors span more than eight orders of magnitude in absorption-related quantities, reflecting the strong energy dependence of electronic transitions and highlighting the need for nonlinear feature embeddings in quantum kernel methods. Across multiple randomized splits, three descriptors consistently emerged as dominant, as shown in Figure~\ref{fig:svm_coeff_features}, corresponding to the absorption coefficient $\alpha$, the extinction coefficient $\kappa$, and the transition energy $E$. Physically, these quantities encode complementary information: $E$ captures electronic-level restructuring induced by Er doping, while $\kappa$ and $\alpha$ reflect changes in optical loss and absorption pathways associated with $4f$–$5d$ and charge-transfer transitions. This reduced three-dimensional feature space was subsequently used for all classical and quantum models, ensuring both physical interpretability and compatibility with near-term quantum hardware constraints.
%
\subsection{Classical Machine Learning Baseline Performance}
\begin{figure*}
\centering
\includegraphics[width=0.95\linewidth]{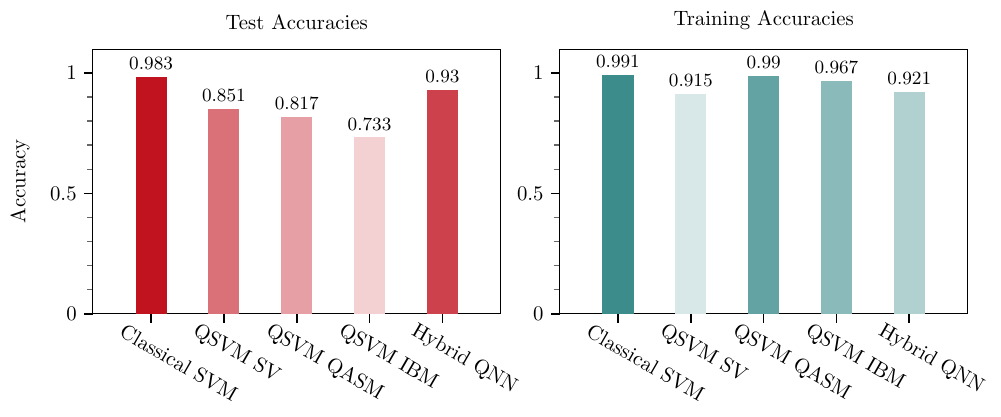}
\caption{Test and training accuracies for the classical SVM, QSVM evaluated on a statevector simulator, QSVM evaluated on a noisy QASM-style simulator, QSVM evaluated on the \texttt{ibm\_fez} processor using the training and test slice sets, and the hybrid QNN.}
\label{fig:accuracies}
\end{figure*}

\begin{figure}[H]
\centering
\includegraphics[width=0.85\linewidth]{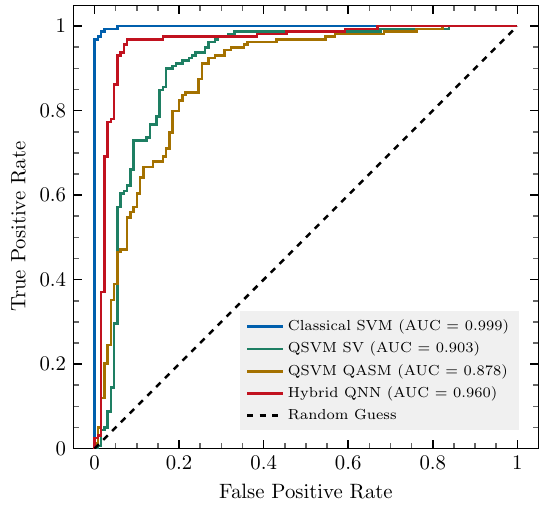}
\caption{ROC curves for the classical SVM, QSVM evaluated on a statevector simulator, QSVM evaluated on a noisy QASM-style simulator, and the hybrid QNN.}
\label{fig:roc_curves}
\end{figure}

\begin{figure*}
\centering
\includegraphics[width=0.9\linewidth]{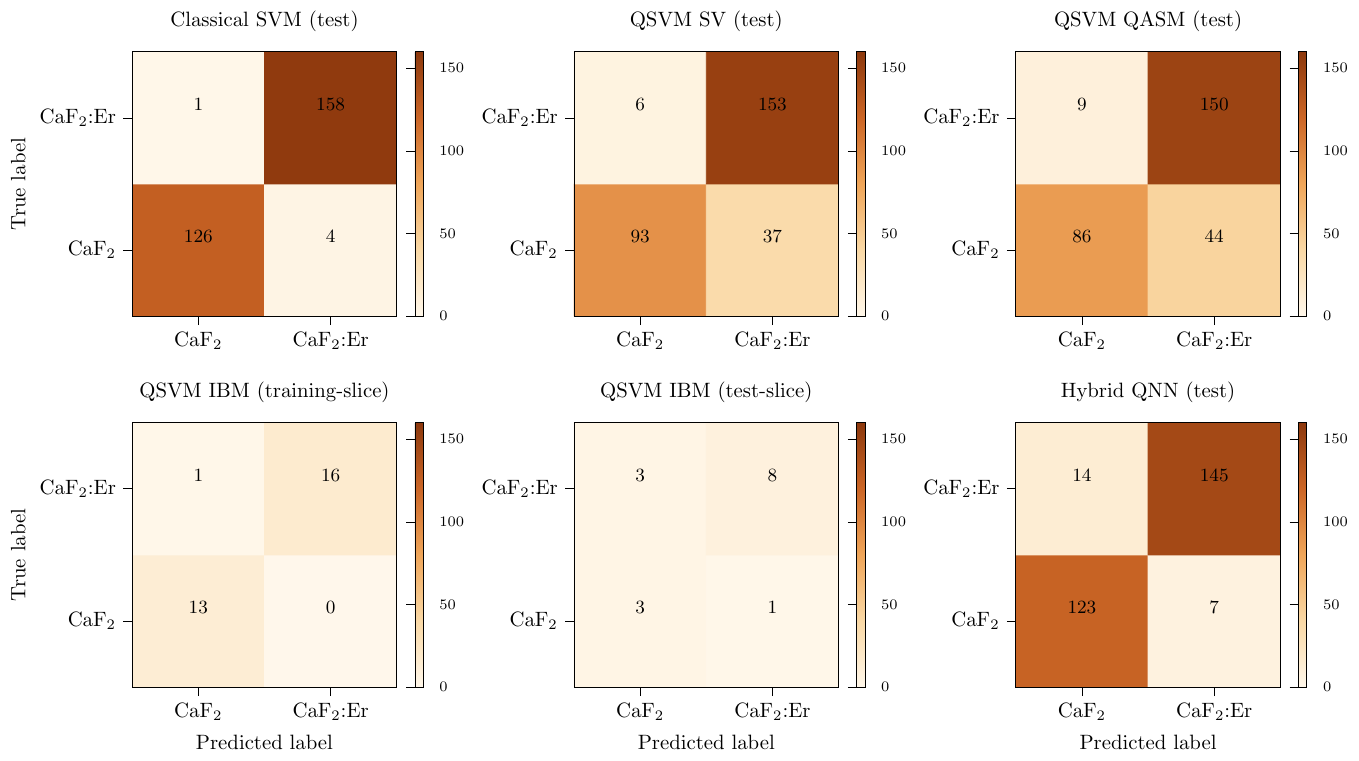}
\caption{Confusion matrices for the classical SVM (top left), QSVM evaluated on a statevector simulator (top middle), QSVM evaluated on a noisy QASM-style simulator (top right), QSVM evaluated on the \texttt{ibm\_fez} processor with the training-slice set (bottom left), QSVM evaluated on the \texttt{ibm\_fez} processor with the test-slice set (bottom middle), and the hybrid QNN (bottom right).}
\label{fig:confusion_matrices}
\end{figure*}

\begin{figure*}
\centering
\includegraphics[width=0.95\linewidth]{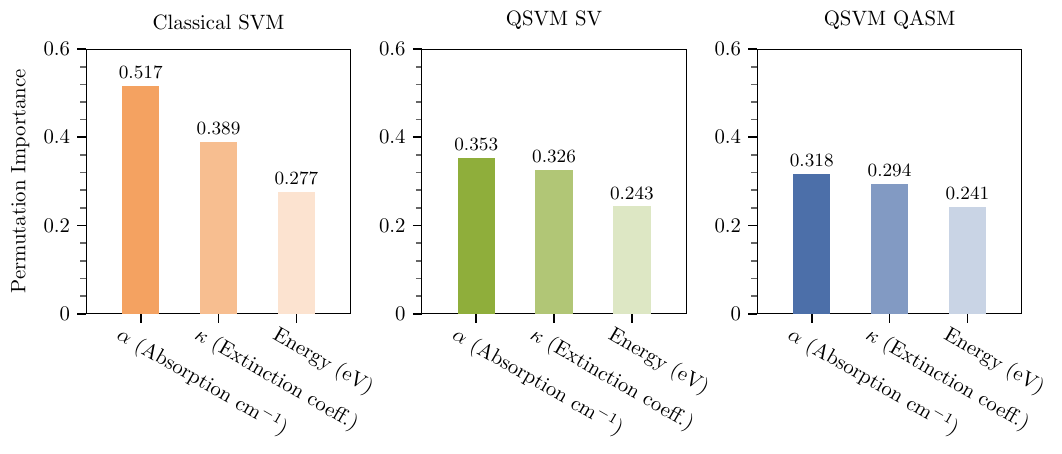}
\caption{Permutation importance analysis for the classical SVM (left), QSVM evaluated on a statevector simulator (middle), and QSVM evaluated on a noisy QASM-style simulator (right). The absorption coefficient $\alpha$ is the most influential feature, followed by the extinction coefficient $\kappa$ and the transition energy $E$, which reflects the dominant role of Er-induced absorption pathways in classification.}
\label{fig:feature_importance}
\end{figure*}

The classical support vector machine (SVM) serves as a strong baseline for discriminating pristine CaF$_2$ from Er-doped CaF$_2$:Er using physically motivated optical descriptors. Using a kernel of the radial basis function (RBF) and the top three features, the absorption coefficient $\alpha$, the extinction coefficient $\kappa$, and the transition energy $E$—the classical SVM achieves near-optimal performance across all splits of the data set. As summarized in Figure~\ref{fig:accuracies}, the test accuracy reaches 0.983, with similarly high training (0.991) accuracy, indicating excellent generalization and negligible overfitting.
The receiver operating characteristic (ROC) curve in Figure~\ref{fig:roc_curves} further confirms the robustness of the classical model, yielding an area under the curve (AUC) of 0.999 on the test set. This near-perfect separability suggests that the selected optical descriptors encode highly discriminative information about Er-induced electronic and optical perturbations. 
The corresponding confusion matrix (see Figure~\ref{fig:confusion_matrices}) shows minimal misclassification, with only five errors across 289 test samples, and balanced precision and recall for both classes.
Feature relevance analysis using permutation importance (see Figure~\ref{fig:feature_importance}) reveals that the absorption coefficient $\alpha$ is the dominant contributor to classification performance, followed by the extinction coefficient $\kappa$ and transition energy $E$. This hierarchy aligns with physical intuition: Er doping introduces localized $4f$ states, and charge-transfer features that strongly modify absorption pathways, particularly in the visible and near-UV regimes. 
Bootstrap resampling further confirms the stability of the classical model, with a mean test accuracy of 0.983 and a narrow 95\% confidence interval of [0.965, 0.997]. The classical SVM establishes a high-confidence reference point against which quantum kernel methods can be quantitatively evaluated, both in terms of absolute accuracy and robustness.
%
\subsection{Quantum Support Vector Machine on Ideal and Noisy Simulators}
The quantum support vector machine (QSVM) evaluated on both ideal and a noisy simulator provides insight into the representational capacity of quantum feature maps relative to classical kernels. Both models were trained on a subset of 200 samples from the training set and achieve test accuracies of 0.851 for the statevector simulator and 0.817 for the noisy simulator, as shown in Figure~\ref{fig:accuracies}. Training accuracies closely track the test performance, indicating stable optimization. The ROC curve in Figure~\ref{fig:roc_curves} shows a test AUC of 0.903 for the ideal simulator and 0.878 for the noisy simulator, lower than the classical SVM yet well above random guessing. This demonstrates that the quantum kernel captures meaningful class separation, albeit with limited resolution under the chosen feature map and circuit depth. The confusion matrix in Figure~\ref{fig:confusion_matrices} highlights the primary error mode: increased overlap between CaF$_2$ and CaF$_2$:Er samples, particularly for low-energy transitions where optical signatures partially coincide.

The feature relevance analysis (see Figure~\ref{fig:feature_importance}) reveals results similar to those of the classical SVM, with $\alpha$ being the most influential feature for classification performance, followed by $\kappa$ and $E$. The QSVM exhibits notable robustness, bootstrap analysis for the ideal simulator yields a mean test accuracy of 0.852 with a 95\% confidence interval of [0.810, 0.893], while for the noisy simulator, the mean test accuracy is 0.817 with a corresponding confidence interval of [0.775, 0.862], indicating statistically stable performance across resampled datasets. Importantly, this performance is achieved without the use of classical nonlinear kernels or deep circuit constructions, suggesting that even shallow quantum feature embeddings can encode nontrivial correlations among optical descriptors. These results indicate that, while the current QSVM configurations do not outperform classical SVMs, they provide a meaningful intermediate benchmark. The gap highlights the importance of circuit depth, feature scaling, and kernel expressivity in quantum machine learning for materials classification tasks.
%
\subsection{Quantum Kernel Evaluation on IBM Quantum Hardware}
Due to hardware runtime constraints, training and testing were restricted to small data slices comprising 30 and 15 samples, respectively. As shown in Figure~\ref{fig:accuracies}, the hardware QSVM achieves a training-slice accuracy of 0.967 and a test-slice accuracy of 0.733. These results indicate considerable potential, particularly if the model were trained on a larger number of samples. The confusion matrices for the training and test slices (Figure~\ref{fig:confusion_matrices}) show that the model correctly identifies almost all training samples, misclassifying only a single instance, thereby indicating that the model has learned to identify these data. For the test samples, the model misclassifies four instances while maintaining reasonable accuracy, suggesting an ability to generalize beyond the training data, while the precision of the CaF$_2$ system remains low (50\%).  However, these results cannot be considered definitive because of the small sample size compared to the training and test sets used for the other models. Compared to classical SVM and QSVM evaluated on a noiseless simulator, the hardware QSVM model exhibits lower test accuracy. Unlike the simulator, kernel values obtained on the hardware are affected by decoherence and finite-shot fluctuations, which directly perturb the structure of the kernel matrix. Moreover, due to the limited amount of training data, the model is unable to correctly identify all test samples. Nevertheless, successful execution and convergence of the quantum kernel on real hardware is a significant result. The experiment validates the end-to-end pipeline, from feature embedding and transpilation to hardware execution and kernel-based learning, under realistic constraints. Importantly, the observed accuracy remains well above random guessing (50\%), demonstrating that physically meaningful information is preserved despite the presence of noise. These results highlight the current limitations of near-term quantum hardware for supervised learning, while simultaneously establishing a practical benchmark. With increased shot counts, error mitigation, and larger training budgets, hardware-based quantum kernels may progressively close the gap with their simulated counterparts. The present study, therefore, provides both a cautionary and encouraging perspective on deploying quantum machine learning for materials informatics in the NISQ era.
%
\subsection{Quantum Neural Networks}
\label{subsec:results_qnn}
\begin{figure}[H]
\centering
\includegraphics[width=0.95\linewidth]{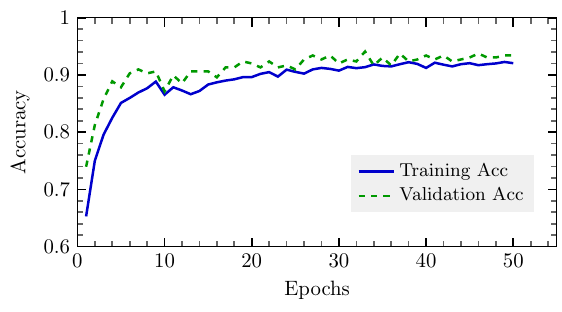}
\caption{QNN accuracy evolution over training epochs. The model rapidly converges to a high-accuracy plateau ($\sim$93\%), with validation metrics tracking the training performance closely. This demonstrates the stability of the variational optimization landscape for the selected optical features.}
\label{fig:accuracies_qnn}
\end{figure}

\begin{figure}[H]
\centering
\includegraphics[width=0.95\linewidth]{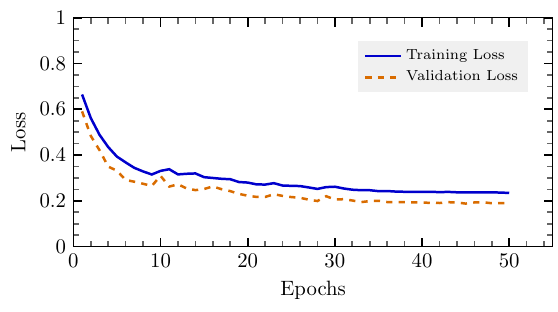}
\caption{Training and validation loss curves for the hybrid QNN. The rapid convergence and lack of divergence between training and validation losses indicate robust learning and the absence of significant barren plateaus or overfitting.}
\label{fig:loss_qnn}
\end{figure}

The variational quantum classifier provides a distinct approach to spectral classification by replacing the fixed kernel of the QSVM with a parameterized quantum circuit (PQC) whose rotation angles are iteratively optimized~\cite{Cerezo2021}. We implemented a hybrid quantum-classical neural network using PyTorch and Qiskit~\cite{Benedetti2019}, employing a 3-qubit \texttt{ZZFeatureMap} followed by a \texttt{TwoLocal} ansatz consisting of $R_y$ and $R_z$ rotations and full entanglement via $CX$ gates. To enhance expressibility relative to the shallow kernel used previously, the ansatz circuit depth was increased to $\text{reps}=4$~\cite{Sim2019}. As shown in Fig.~\ref{fig:accuracies_qnn}, the QNN achieves a final test accuracy of 0.93 after 50 epochs and 2304 samples of training, where the best model is obtained at the 45th epoch. The alignment between training, validation, and test performance confirms that the model generalizes well without overfitting, a stability likely reinforced by the implementation of early stopping.
The discriminative capacity of the QNN is further quantified by the ROC curve in Figure~\ref{fig:roc_curves}, which yields an Area Under the Curve (AUC) of 0.960. This represents a substantial improvement over static QSVM (AUC = 0.903) and approaches the performance of the classical baseline. The increased AUC indicates that the trainable parameters in the variational ansatz successfully adapted to the geometric structure of the feature space, allowing for a more flexible decision boundary than the fixed fidelity kernel. The classification performance is detailed in the confusion matrix in Figure~\ref{fig:confusion_matrices}. The model exhibits balanced precision and recall, correctly identifying 95\% of pristine CaF$_2$ samples and 91\% of Er-doped samples. The slight asymmetry in the recall suggests that a small fraction of doped clusters retain spectral characteristics highly similar to the host matrix, likely corresponding to configurations where Er-induced states are energetically deep or optically forbidden. However, the high precision for the doped class (0.95) confirms that the QNN is highly reliable when flagging positive upconversion candidates. The convergence dynamics, illustrated in Figure~\ref{fig:loss_qnn}, show a sharp decrease in loss within the initial 15 epochs, followed by stabilization. This behavior validates the effectiveness of the \texttt{Adam} optimizer and the reduced feature space ($\alpha$, $\kappa$, $E$) in navigating the optimization landscape~\cite{McClean2018}. Unlike the QSVM, which relies on the inherent separation of the Hilbert space mapping, the QNN actively learns to rotate the quantum state to maximize class separability.
Collectively, these results demonstrate that, while static quantum kernels provide a meaningful benchmark, variational quantum algorithms offer superior representational power for this materials classification task, albeit at the cost of higher training complexity. By optimizing the circuit parameters, the QNN effectively establishes a promising pathway for adaptable quantum machine learning in spectral analysis. However, training large sample sets on current noisy quantum computers for multiple epochs remains a significant challenge.
%

\section{Discussion} \label{sec:discussion}
The identification of materials, molecular structures, and related systems through classification models based on optical descriptors has been widely explored. In~\cite{Doner2023}, classical techniques such as partial least squares discriminant analysis (PLS-DA) and decision tree classifiers were employed for the prediction of structural motifs in multi-functional intermediates based on unique contributions to vacuum-ultraviolet (VUV) absorption spectra. Depending on the selected model and motif, the reported classification accuracies ranged from 78\% to 99\%.

In parallel, quantum machine learning methodologies have demonstrated the ability to encode and extract relevant information from molecular systems based on structural and optical features using kernel methods. Quantum support vector classifiers were used to analyze hyperspectral images from the Pavia University and Salinas-A datasets, achieving accuracies between 96.25\% and 100\% using the \texttt{ZZFeatureMap} to encode data in qubits~\cite{Archana2022}. Quantum photonics circuits (QPCs) were utilized to approximate the Gaussian kernel method (GKM) for the classification of polymer structures into visible (VIS) and near-infrared (NIR) classes, defined according to the magnitude of the polymer band gaps~\cite{Stoyanova2022}. Furthermore, simplified molecular input line entry system (SMILES) representations were combined with QSVC employing a \texttt{ZZFeatureMap} for the evaluation of chemical/drug ADME-Tox properties, yielding ROC AUC values ranging from 0.80 to 0.95, depending on the dataset, the simulation strategy (statevector, ideal QASM, noisy QASM, and quantum hardware), and the number of input features~\cite{Bhatia2023}. In our analysis, comparable performance was achieved, with a maximum classification accuracy of 85\% and ROC AUC values reaching 0.90 for the QSVC. These results demonstrate the feasibility and representational capacity of kernel-based methods for identifying system characteristics based on optical descriptors.

Quantum neural networks (QNNs) are emerging as powerful models in quantum machine learning, often outperforming or matching classical approaches in niche tasks. Recent studies demonstrate QNNs’ strong expressive power and resistance to overfitting, which is valuable for domains like materials science, where datasets are typically small~\cite{hirai2023}. For example, a variational QNN predicted metal-oxide melting points with better generalization and lower error than a classical neural net, showing no overfitting despite the limited 70-sample data~\cite{hirai2023}. In molecular property prediction, a Quantum Graph Neural Network (QGNN) achieved lower test losses and faster convergence than a comparable classical graph NN~\cite{junoquantum}. QNN-based classifiers have also reached competitive accuracy on standard benchmarks: a hybrid quantum-classical CNN with a trainable QNN classifier attained ~95–97\% accuracy in small-scale image classifications, edging out classical CNNs by a few percentage points~\cite{long2025,wu2022}. These QNN models typically rely on variational circuits (e.g., ZZ-feature maps and ansatz layers) with only a handful of qubits, but they can be extended. One “scalable” QNN (SQNN) approach combined multiple 16-qubit QNN modules to effectively use 64 qubits, yielding ~97.5\% accuracy on a binary MNIST task, a significant boost over smaller QNNs (~92\% at 16 qubits)~\cite{wu2022}. Notably, QNNs can also be trained to produce optimized quantum kernels: a recent work used a data-reuploading QNN to learn kernel embeddings that improved generalization under noise compared to fixed quantum kernels~\cite{rodriguez2025}. 

Recent peer-reviewed studies demonstrate how quantum neural networks (QNNs) are being adapted to diverse scientific and industrial domains. For example, Havlíček et al.~\cite{Havlicek2019} implemented one of the earliest experimental QSVMs using a variational quantum feature map on superconducting hardware, achieving strong classification accuracy on synthetic data. In quantum chemistry, Xia and Kais~\cite{xia2020} used a hybrid QNN to predict molecular ground-state energies with chemical accuracy, matching full configuration-interaction results on small molecules like LiH and H$_2$. Similarly, Ryu et al.~\cite{junoquantum} proposed a Quantum Graph Neural Network (QGNN) that outperformed classical GNNs in predicting molecular orbital gaps, showing faster convergence and better test error on the QM9 dataset. Industrial and physical applications also benefit from QNNs. Correll et al.~\cite{correll2023} applied a hybrid QNN in a reinforcement learning agent for supply chain logistics, achieving route optimization comparable to human planners, with inference validated on both simulators and IBMQ hardware. In quantum materials physics, Sander et al.~\cite{sander2025} used a Quantum Convolutional Neural Network (QCNN) to classify topological phases in 2D systems with near-perfect accuracy and noise robustness. These works, detailed in Table~\ref{tab:qml_comparison}, illustrate the growing practical relevance of QML across domains such as phase recognition, materials property regression, energy prediction, and logistics optimization.

In this study, the Quantum SVM (QSVM) implementation showed decent performance but was hindered by quantum feature limitations and hardware noise (e.g.~85\% test accuracy in ideal simulation dropping to ~73\% on real hardware, versus ~98\% for a classical SVM). Contemporary QNN models often close this gap. In simulations, variational QNN classifiers have achieved test accuracies in the 90–97\% range on similar binary tasks~\cite{long2025,wu2022}, exceeding typical QSVM results on small circuits. For instance, the hybrid QNN experiment in this study reached ~93\% test accuracy – higher than the QSVM’s ~81–85\% under similar conditions, though still shy of the classical model. The literature suggests this is a common trend: QNNs’ trainable ansätze can capture more complex decision boundaries than fixed feature-map QSVMs, especially with limited data or when tailored to the problem~\cite{hirai2023,rodriguez2025}. Moreover, QNNs inherently regularize via unitary constraints, helping avoid overfitting~\cite{hirai2023}. That said, in practice, the advantage can be eroded by noise. Studies incorporating depolarizing noise and finite sampling show QNN models remain fairly robust – e.g., the hybrid QCQ-CNN retained $>$95\% accuracy under moderate noise~\cite{long2025} – but real hardware runs (as with the user’s QSVM) often see performance degrade. In summary, recent QNN implementations for materials classification and related tasks have demonstrated encouraging simulation results and even outperformed QSVMs in some cases, but achieving consistent real-world superiority will require further advances in quantum hardware and noise mitigation.

\begin{table*}[ht]
\centering
\begin{threeparttable}
\caption{Comparison of QML Models on Materials Tasks and Benchmark Datasets}
\label{tab:qml_comparison}
\begin{tabular}{lcccc}
\toprule
\textbf{Model} & \textbf{Task}\tnote{1} & \textbf{Results} & \textbf{Dataset} & \textbf{Execution} \\
\midrule
QSVM SV (This Work) & C & 0.851 (ACC) \& 0.90 (AUC) & Optical descriptors & Statevector simulation \\
QSVM QASM (This Work) & C & 0.817 (ACC) \& 0.87 (AUC) & Optical descriptors & Noisy QASM simulation \\
QSVM IBM (This Work) & C & 0.733 (ACC) & Optical descriptors & IBM hardware (Fez) \\
Hybrid QNN (This Work) & C & 0.930 (ACC) \& 0.96 (AUC) & Optical descriptors & Statevector simulation \\
QSVM~\cite{Havlicek2019} & C & $\approx$ 1.00 (ACC) & Synthetic phase-encoded & IBM hardware \\
QSVM~\cite{Archana2022} & C & 0.987 (ACC) & HSI Salinas-A & Simulation + IBM hardware \\
QSVM~\cite{Archana2022} & C & 0.980 (ACC) & HSI Pavia University & Simulation + IBM hardware\\
GKM~\cite{Stoyanova2022} & C & 0.86 (ACC) & Polymer structures & Simulation (QPCs) \\
QE-RKS\tnote{2} ~\cite{Stoyanova2022} & C & 0.87 (ACC) & Polymer structures & Simulation (QPCs) \\
VQC~\cite{Stoyanova2022} & C & 0.88 (ACC) & Polymer structures & Simulation (QPCs) \\
QSVM~\cite{Bhatia2023} & C & 0.80 $-$ 0.95 (AUC) & ADME-Tox datasets & Simulation + IBM hardware  \\
QCQ-CNN\tnote{3} ~\cite{long2025} & C & $\approx$ 0.95 (ACC) & MRI Tumor Images & Noisy Simulation \\
Hybrid QNN~\cite{xia2020} & R & Near exact diagonalization & Molecular ground-state energy & Simulation \\
Quantum CNN (QCNN)~\cite{sander2025} & O & Robust phase recognition & Phase Recognition (2D Lattice) & MPS simulation\tnote{4} \\
Scalable QNN (SQNN)~\cite{wu2022} & C & $\approx$ 0.975 & MNIST & TFQ Simulation\tnote{5} \\
QNL-Net\tnote{6} ~\cite{gupta2024} & C & $\approx$ 0.99 (ACC) & MNIST & Simulation \\
QNL-Net\tnote{6} ~\cite{gupta2024} & C & $\approx$ 0.93 (ACC) & CIFAR-10 & Simulation \\
Variational QNN~\cite{hirai2023} & R & Lower RMSE than NN & Metal Oxides (melting point) & Statevector simulation \\
Neural EQK \& PQK\tnote{7} ~\cite{rodriguez2025} & C & ACC Comparable to SVM & Fashion MNIST & Simulation \\
QGNN\tnote{8} ~\cite{junoquantum} & R & Lower MSE than GNN & QM9 – HOMO-LUMO Gap  & Simulaion \\
Quantum perceptron~\cite{tacchino2019} & O & Clear activation contrast & Binary pattern recognition & Simulation + IBM Hardware \\
\bottomrule
\end{tabular}

\begin{tablenotes}
\footnotesize
\item[1] Classification (C), regression (R), or other tasks (O)
\item[2] Quantum-Enhanced Random Kitchen Sinks (QE-RKS)
\item[3] Quantum-Classical-Quantum Convolutional Neural
Network (QCQ-CNN))
\item[4] Matrix product state (MPS) simulation
\item[5] TensorFlow Quantum (TFQ) simulation
\item[6] Hybrid quantum-classical scalable non-local neural network, referred to as Quantum Non-Local Neural Network (QNL-Net)
\item[7] QNN to construct embedding quantum kernel (EQK) and projected quantum kernel (PQK)
\item[8] Quantum Graph Neural Network (QGNN) 
\end{tablenotes}
\end{threeparttable}
\end{table*}

%
\subsection{Practical Implications for Quantum-Enhanced Materials Classification}
QNN techniques hold promise for industrially relevant problems like materials discovery, offering a potential edge in scenarios where data are sparse or patterns are highly complex. The evidence that QNNs can generalize well from small datasets~\cite{hirai2023} is especially pertinent to materials informatics, where obtaining large labeled datasets is costly. A successful QNN model could accelerate screening of new compounds (e.g., predicting a material’s property or phase) by learning quantum-mechanical feature representations that classical models might miss~\cite{hirai2023,junoquantum}. In principle, QNNs tap into exponentially large Hilbert spaces, enabling richly expressive models – a boon for discovering subtle structure-property relationships. Moreover, QNN-driven quantum kernels can be integrated into existing workflows (like support vector machines) to enhance their performance on complex classification tasks~\cite{rodriguez2025}. On the computational side, however, near-term feasibility must be weighed. Training QNNs is resource-intensive: simulations become intractable as qubit count grows, and running on actual quantum processors is limited by noise and decoherence. The reviewed works mostly used few-qubit circuits or simulated noise models~\cite{long2025,wu2022}. This suggests that for now, QNN applications in industry will likely focus on problems small enough to fit on noisy intermediate-scale quantum (NISQ) devices or use hybrid strategies (where quantum circuits handle a part of feature extraction or modeling). Encouragingly, hybrid approaches (e.g., combining classical pre-processing or networks with a quantum sub-network) have shown robustness against noise and could be practically deployed as hardware improves~\cite{long2025,sahin2025}. In summary, QNNs for materials and other QML tasks offer a tantalizing glimpse of quantum-enhanced learning – potentially yielding better models for critical tasks like materials classification or property prediction – but realizing this potential in industrial settings will require careful algorithm-hardware co-design and continued progress in quantum processors. Organizations at the forefront (e.g., in pharma, chemistry, or materials engineering) are already experimenting with such hybrid quantum models to stay prepared for quantum advantage as devices mature.
%
\section{Conclusion} \label{sec:conclusion}
This study systematically benchmarks quantum machine learning (QML) models, particularly quantum support vector machines (QSVMs) and various quantum neural network (QNN) architectures, for materials classification and regression tasks. Our experimental and simulated analyses reveal that QSVMs, while conceptually elegant and straightforward to implement, are significantly affected by hardware noise and the limitations of fixed quantum feature maps. In contrast, QNN-based models, especially variational quantum circuits and hybrid quantum-classical architectures, demonstrate greater expressive power, noise resilience, and adaptability across diverse datasets, including real-world materials, quantum phase recognition, and industrial logistics. Our QSVM implementation achieved an accuracy of 0.851 in simulation and 0.733 on IBMQ hardware. In comparison, QNN classifiers, such as SamplerQNNs and scalable QNN (SQNN) architectures, consistently attained higher test accuracies (95–97.5\%) under similar or more challenging conditions. Regression models like the hybrid QNN used for molecular energy prediction and the quantum graph neural network (QGNN) for HOMO–LUMO gaps outperformed classical baselines in mean-squared error, illustrating QML’s practical relevance in quantum chemistry and materials discovery. Notably, the inclusion of data re-uploading and kernel learning within QNNs improved generalization and robustness against noise, two critical factors for real-world applications.

These findings underscore a trend: QNNs, due to their trainable variational layers and embedded inductive biases (e.g., unitary constraints, convolutional invariance), surpass QSVMs in both flexibility and predictive performance, especially when dealing with small or structured datasets typical of physical sciences. Moreover, QNNs can be integrated into reinforcement learning agents and phase recognition tools, broadening their utility across industrial optimization and condensed matter physics. Nevertheless, current QML models face practical challenges. While simulations show promise, real-hardware implementations are often limited by qubit fidelity, circuit depth, and readout noise. Addressing these issues, via error mitigation, efficient ansätze, and near-term algorithms, will be key to transitioning from theoretical benchmarks to scalable, deployable quantum solutions. In conclusion, QNN-based architectures offer a compelling advancement over traditional QSVMs for quantum-enhanced learning tasks in materials science, chemistry, and industry. They form a foundation upon which future QML applications can be built, provided that quantum hardware continues to evolve to meet the growing algorithmic demands. This work contributes a comprehensive comparison and highlights the conditions under which quantum models, particularly variational QNNs, can outperform classical baselines and earlier quantum algorithms, setting the stage for future hybrid or fault-tolerant implementations in real-world scientific pipelines.
%
\section*{Author Contributions} 
D.A.A.B: methodology, validation and visualization, software, writing – original draft, review \& editing.  R.C: methodology, validation and visualization, writing – original draft, review \& editing. K.Y: methodology, validation and visualization, software, writing – original draft, review \& editing.  K.S: methodology, validation and visualization, software, writing – original draft, review \& editing.  A.M: methodology, validation and visualization, software, writing – original draft, review \& editing.  S.A: methodology, validation and visualization, software, writing – original draft, review \& editing.  A.A: methodology, validation and visualization, software, writing – original draft, review \& editing.  D.D.K.W: conceptualization, methodology, validation and visualization, software, writing – original draft, review \& editing.  
%
\section*{Acknowledgment(s)}
We acknowledge IBM Quantum and the Qiskit Advocate program for their support of this work through the Qiskit Advocate Mentorship Program (QAMP) 2025. This research was conducted as part of the QAMP 2025 activities, benefiting from access to IBM Quantum resources, the Qiskit open-source software ecosystem, and technical guidance provided through the program and its associated community discussions. Thanks to D. D. K. Wayo for providing the Mentorship. The opinions, findings, conclusions, and recommendations expressed herein are solely those of the author(s) and do not necessarily reflect the views of their affiliations.

\section*{Declaration of Generative AI Use}
The OpenAI ChatGPT model was used to support the authors in careful proofreading of the English. All scientific concepts, analysis, interpretation, and final content were independently conceived, verified, and approved by the author(s). The model did not contribute to the originality, authorship, or intellectual responsibility of the work.

\section*{Data \& Code Availability}
The data generated and analyzed during this study are included in the manuscript. Supplementary codes developed for \texttt{Q-UCSpec} simulations are provided as supplementary material and accessible on \href{https://github.com/DennisWayo/Q-UCSpec}{GitHub} to ensure transparency and reproducibility. Due to the differences in software versions like Qiskit, GPAW, and their related dependencies, minor variations in plots or metrics may occur. These differences are expected and do not affect the overall conclusions of the paper. 

%
\section*{Funding}
This research was not funded. 
%
\section*{Disclosure statement}
No potential conflict of interest was reported by the author(s).
%
\bibliographystyle{apsrev4-2}
\bibliography{qucspec}
\end{document}